\documentclass[12pt,final]{elsarticle}
\usepackage{latexsym}
\usepackage{amsfonts}
\usepackage[latin1]{inputenc}
\usepackage{subfig}
\usepackage{graphicx}
\usepackage{mathrsfs}
\usepackage{amssymb}
\usepackage{amsmath}
\usepackage{verbatim}
\usepackage[dvips]{color}

\begin{document}
\begin{frontmatter}
\title{Minimal $Z'$ models and the $125$ GeV Higgs boson}
\author{L.~Basso}
\address{Physikalisches Institut, Albert-Ludwigs-Universit\"at Freiburg
D-79104 Freiburg, Germany}
\ead{lorenzo.basso@physik.uni-freiburg.de}

\begin{abstract}
The $1$-loop renormalization group equations for the minimal $Z'$ models encompassing a type-I seesaw mechanism are studied in the light of the $125$ GeV Higgs boson observation. This model is taken as a benchmark for the general case of singlet extensions of the standard model. The most important result is that negative scalar mixing angles are favored with respect to positive values. Further, a minimum value for the latter exists, as well as a maximum value for the masses of the heavy neutrinos, depending on the vacuum expectation value of the singlet scalar.

\end{abstract}
\end{frontmatter}

\section{Introduction}
\label{sect:intro}
Both the ATLAS and CMS Collaborations have reported the discovery of a resonance
with a mass of around $125$ GeV, immediately associated with the long-sought after Higgs boson~\cite{Observation}.
With this discovery, the standard model (SM) of particle physics is considered to be complete. However, such low mass is incompatible~\footnote{This tension can be softened if one relaxes the requirement of stability of the vacuum up to the Planck scale to metastability with lifetime larger than the age of the universe, that is a sufficient requirement for stability. Barring this in mind, in this Letter we will consider only the former because it is more restrictive.} with the stability of the SM up to the Planck scale ($M_{Pl}$), as newly confirmed by a 2-loop analysis of the scalar potential~\cite{Degrassi:2012ry} (see, however,~\cite{Alekhin:2012py,Masina:2012tz}).

It is also known that simple extensions of the SM can ameliorate this behavior. It is sufficient to enlarge the scalar sector by means of a singlet scalar field to restore the validity of the SM up to very large scales (see, e.g., \cite{Extra_scalars,Bowen:2007ia} and references therein). Nonetheless, other problems 
are still unsolved, for instance the pattern of masses and mixing angles of the neutrinos and the presence of dark matter
 just to mention a few. Extensions in more SM sectors than only in the scalar one are therefore needed.

A very simple and yet interesting SM extension is the so-called ``minimal $Z'$ models'', in which the SM gauge group is enlarged by a $U(1)_{B-L}$ gauged symmetry~\cite{minimal_Zp}. The vanishing of chiral anomalies requires to minimally include $3$ right-handed (RH) neutrinos, while the extra $U(1)$ symmetry is spontaneously broken by a new complex singlet scalar field, responsible for giving mass to the related $Z'$ boson. While the scalar and gauge sectors are unambiguous, various realizations exist for the fermion content of the model, that can encompass a gauged type-I seesaw mechanism in its minimal version~\cite{minimal_Zp_typeI,Basso:2010jm}, or the more involved inverse seesaw mechanism if new lepton pairs are included, thereby providing a solution for neutrino masses, dark matter and baryon asymmetry at once~\cite{Basso:2012ti}. Higher fermion representations can be invoked to explain the apparent enhancement in the Higgs-to-diphoton signal as recently observed, if one consider negative values for the scalar mixing angle, compatible with LEP and LHC observations, as in Ref.~\cite{Basso:2012nh}.

It is a fact that the evolution of the scalar parameters in simple SM extensions does depend only very marginally on the gauge sector and sensibly on the fermions only if the latter are very heavy. This has explicitly been noted in Ref.~\cite{Bowen:2007ia}, although this reference does not study in detail its consequences. Further, in Ref.~\cite{Basso:2010jm} a renormalization group (RG) study of the minimal $Z'$ models has been performed when the SM-like Higgs mass was not known. Stimulated by these arguments, in this Letter we study the constraints that the $125$ GeV SM-like Higgs boson imposes on the minimal $Z'$ models via the 1-loop RG running in the latter, taken as a benchmark for minimal scalar extensions of the SM. Imposing stability up to the Planck scale, we are able to determine predictions for the mass of the heavy scalar field and constraints on the scalar mixing angle. In a model-dependent analysis, we will show the existence of a minimum value for the latter and of an upper value for the mass of the heavy neutrinos, when considering just the minimal case of the type-I seesaw mechanism. For a similar analysis in the SM with RH neutrinos and an effective type-I seesaw mechanism, see~\cite{Masina:2012tz}. We will discuss how all these statements depend upon the $U(1)_{B-L}$-breaking scalar vacuum expectation value (vev).

\section{The minimal $Z'$ models}
\label{sect:model}
We describe here the minimal $Z'$ models. Following the notation of Ref.~\cite{Basso:2010jm}, the SM gauge group is extended by a gauged $U(1)_{B-L}$ symmetry, such that the classical gauge invariant Lagrangian obeys the $SU(3)_C\times SU(2)_L\times U(1)_Y\times U(1)_{B-L}$ gauge group.

When multiple $U(1)$ factors are present, the related gauge fields tend to mix.
In a suitable basis with diagonal kinetic terms, the mixing is shifted to the covariant derivative. For the neutral sector, this reads
\begin{equation}\label{cov_der}
D_{\mu}\equiv \partial _{\mu} + \dots  +ig_1YB_{\mu} +i(\widetilde{g}Y + g_1'Y_{B-L})B'_{\mu}\, .
\end{equation}
The last term of Eq.~(\ref{cov_der}) can be rewritten defining an effective coupling $Y^E$ and an effective charge $g_E$, thus $g_E Y^E \equiv \, \widetilde{g}Y + g_1'Y_{B-L}$~\footnote{Some authors prefer to identify the $U(1)$ factors appearing in the gauge group with the charges that the $Z'$ boson directly couples to ($Y$ and $Y^E$, see, e.g.,~\cite{minimal_Zp}). Hence, the notation $SM\times U(1)_{B-L}$ might generate some confusion. In the following, we always consider the case of generic mixing between the $U(1)_Y$ and $U(1)_{B-L}$ gauge groups, where the new $Z'$ boson couples to one such linear combination. The model in which the $Z'$ boson couples only to the $B-L$ charge, corresponding to $\widetilde{g}=0$, is called `pure $B-L$ model' and represents only one specific benchmark model of our generic parameterization.}.
As any other parameter in the Lagrangian, $\widetilde{g}$ and $g_1'$ are running parameters, therefore their values ought to be defined at some scale~\cite{running_U(1)}. A discrete set of popular $Z'$ models can be recovered by a suitable definition of both $\widetilde{g}$ and $g_1'$. (For more details, see~\cite{Basso:2010jm} and references therein.)

A new complex scalar field $\chi$ is required to provide mass to the further gauge boson. The scalar potential is then given by
\begin{equation}\label{BL-potential}
V(H,\chi ) = m^2H^{\dagger}H + \mu ^2\mid\chi\mid ^2 + \lambda _1 (H^{\dagger}H)^2 +\lambda _2 \mid\chi\mid ^4 + \lambda _3 H^{\dagger}H\mid\chi\mid ^2  \, ,
\end{equation}
where $H$ and $\chi$ are the complex scalar Higgs doublet and singlet fields with $B-L$ charges $0$ and $+2$, respectively. In unitary gauge, we parametrize $
H \equiv \left( \begin{array}{c} 0 \\ \frac{v+h}{\sqrt{2}} \end{array} \right)$ and $\displaystyle \chi \equiv \frac{x+h'}{\sqrt{2}}$
with $v$ and $x$ real and non-negative.  We denote by
$h_1$ and $h_2$ the scalar fields of definite masses, $m_{h_1}$ and $m_{h_2}$ respectively, and we conventionally choose
$m^2_{h_1} < m^2_{h_2}$. Further, we call $\alpha$ the mixing angle between the real scalar states
\begin{equation}\label{displ_scalari_autostati_massa}
\left( \begin{array}{c} h_1\\h_2\end{array}\right) = \left( \begin{array}{cc} \cos{\alpha}&-\sin{\alpha}\\ \sin{\alpha}&\cos{\alpha}
	\end{array}\right) \left( \begin{array}{c} h\\h'\end{array}\right) \, ,
\end{equation}
 with $-\frac{\pi}{2}\leq \alpha \leq \frac{\pi}{2}$.
For our numerical study, it is useful to rewrite the parameters in the Lagrangian in terms of the physical quantities $m_{h_1}$, $m_{h_2}$ and $\sin{2\alpha}$ after inversion:
\begin{eqnarray}\nonumber
\lambda _1 &=& \frac{m_{h_1}^2}{4v^2}(1+\cos{2\alpha}) + \frac{m_{h_2}^2}{4v^2}(1-\cos{2\alpha}),\\ \nonumber
\lambda _2 &=& \frac{m_{h_2}^2}{4x^2}(1+\cos{2\alpha}) + \frac{m_{h_1}^2}{4x^2}(1-\cos{2\alpha}),\\ \label{inversion}
\lambda _3 &=& \sin{2\alpha} \left( \frac{m_{h_2}^2-m_{h_1}^2}{2xv} \right).
\end{eqnarray}

Generally, a bound on $x$ is set by LEP measurements. In the type-I seesaw case here discussed, $x > 3.5$ TeV~\cite{Basso:2010jm}. The latest LHC data~\cite{New_Higgs_data} are interpreted by identifying $h_1$ with the recently observed boson, i.e., $m_{h_1}=(125\pm 1)$ GeV, and by restricting $|\sin{\alpha}| \leq 0.3$. The latter value is compatible with the observed slight reduction of the Higgs tree-level decays, which however still suffer of large uncertainties. 
Finally, the influence of the extra scalar on the SM quartic coupling is to uplift it by a positive quantity given by
\begin{equation}\label{delta_lambda}
\lambda_1 - \lambda^{SM}_1 \equiv \Delta \lambda_1 = \frac{(\sin{\alpha})^2}{2 v^2} (m^2_{h_2}-m^2_{h_1}),
\end{equation}
where $\lambda^{SM} = (m^2_h)_{SM}/(2 v^2)$ is the SM value.

We have so far described the sectors that all the various realizations of the minimal $Z'$ models have in common. As a representative for the fermion sector, we consider here the type-I seesaw mechanism, that requires to introduce $1$ RH neutrino per generation ($\nu_R$, with $Y^{\nu_R}_{B-L}=-1$) to ensure the vanishing of the chiral anomalies. The new Yukawa interactions are
\begin{equation} \label{L_Yukawa}
\mathscr{L}^\nu_Y = -y^{\nu}_{jk}\overline {l_{jL}} \nu _{kR}\widetilde H 
	         -y^M_{jk}\overline {(\nu _R)^c_j} \nu _{kR}\chi +  {\rm 
h.c.}  \, ,
\end{equation}
{where $\widetilde H=i\sigma^2 H^*$ and  $i,j,k$ take the values $1$ to $3$},
where the last term is the Majorana contribution and the first term the usual Dirac one. These are the only allowed gauge invariant terms, provided that $Y^\chi_{B-L} = 2$. Neutrino mass eigenstates, obtained after applying the seesaw mechanism, will be called $\nu_l$ (with $l$ standing for light) and $\nu_h$ (with $h$ standing for heavy), where the first ones are the SM-like ones.

The complete set of RGEs for the generic model are derived for the parameters in the Lagrangian and can be found in Ref.~\cite{Basso:2010jm}~\footnote{Compared to Ref.~\cite{Basso:2010jm}, the correct coefficient of the $(y^M)^4$ term in Eq.~(\ref{RGE_lambda2}) is $-16$ and not $-1$. See Ref.~\cite{Corrected_RGEs}.}. We show here only those pertaining to the scalar quartic couplings, subject of this Letter:
\begin{eqnarray}\label{RGE_lambda1}
\beta _1 \equiv \frac{d \lambda _1}{dt} &=&  \frac{1}{16\pi ^2}\left( 24\lambda _1^2+\lambda _3^2 -6y_t^4 + 12\lambda _1 y_t^2 \right)\, ,\\ \label{RGE_lambda2}
\beta _2 \equiv \frac{d \lambda _2}{dt} &=&  \frac{1}{8\pi ^2}\left( 10\lambda _2^2+\lambda _3^2-8Tr\left[ (y^M)^4\right] +4\lambda _2Tr\left[ (y^M)^2\right] \right)\, ,\\ \label{RGE_lambda3} 
\beta _3 \equiv \frac{d \lambda _3}{dt} &=&  \frac{\lambda _3}{8\pi ^2}\left( 6\lambda _1+4\lambda _2+2\lambda _3+3y_t^2 +2Tr\left[ (y^M)^2\right]\right)\, ,
\end{eqnarray}
with $t=\ln{Q}$. Notice that sub-leading gauge coupling terms are not shown because they are irrelevant for the discussion, but they have been taken into account in the derivation of the results. These equations match those in Ref.~\cite{Bowen:2007ia}, and make explicit the model-dependent contribution arising from the fermionic sector. When the neutrinos are relatively light and their Yukawa couplings small (meaning that their contribution can be neglected), our results can be interpreted as general for the class of singlet-extended SM theories. (For similar analyses of the SM with singlet extensions, see~\cite{singlet} and references therein.)

For their numerical study, we proceed as in~\cite{Basso:2010jm}, putting boundary conditions at the electroweak (EW) scale on the free physical observables: $m_{h_2}$, $\alpha$, $x$, $M^{1,2,3}_{\nu_h}$, that we trade for $\lambda _1$, $\lambda _2$, $\lambda _3$, $y^M_{1,2,3}$. The impact of the variation of the Higgs and top masses within $1\sigma$ is also considered, being $m_{h_1} = (125\pm 1)$ GeV and $m_{top} = (173.1 \pm 0.7)$ GeV, respectively. Finally, gauge parameters are sub-leading in our considerations and need not be specified. For completeness, we used the same values as in~\cite{Basso:2010jm}. The free parameters in our study are then $m_{h_2}$ (the heavy Higgs mass), $\alpha$ (the scalar mixing angle), $x$ (the ($B-L$)-breaking vev) and $M_{\nu _h}$ (the heavy neutrino mass when taking neutrinos as mass degenerate). The general philosophy is to fix in turn some of the free parameters and scan over the other ones, individuating the allowed regions fulfilling the following set of conditions:
\begin{equation}\label{cond_1}
\lambda _{1,2,|3|}(Q) < 1 \qquad \forall \; Q \leq Q'\, ,
\end{equation}
(usually referred to as the ``triviality'' condition. Notice that $\lambda _{|3|}\equiv |\lambda _3|$) and
\begin{equation}\label{cond_2}
0 < \lambda _{1,2}(Q) \qquad \mbox{and} \qquad
4\lambda _1(Q)\lambda _2(Q)-\lambda _{3}^2(Q) > 0 \qquad \forall \; Q \leq Q'\, ,
\end{equation}
(usually referred to as the ``vacuum stability'' condition). The meaning of Eqs.~(\ref{cond_1})--(\ref{cond_2}) is that the request of perturbativity and the well-definition of the vacuum of the model must be fulfilled at any scale $Q\leq Q'$, being $Q'$ the ultimate scale of validity of the theory. In this Letter, we take $Q'=M_{Pl}=10^{19}$ GeV unless otherwise specified. 
In these equations,
we have defined a parameter to be ``perturbative'' for values less than unity. 
Both conditions have been chosen to be most restrictive to show that the model under consideration is a simple SM extension that is viable up to the Planck scale and no further completion is required, and to identify the tightest possible bounds in parameter space. Relaxing these conditions (e.g., considering $Q'<M_{Pl}$ or increasing the perturbativity bound) will lead to results of which our findings are a still valid subset.
Regarding Eq.~(\ref{cond_2}) and in particular its second term, notice that, in contrast to the SM in which it is sufficient the Higgs self-coupling $\lambda$ be positive, in this model said condition can be violated even for positive $\lambda _{1,2,3}$.

\section{Numerical Results}
\label{sect:results}
To understand the following results, some considerations on the RGEs are in place. First, the reason why extending the SM by a singlet scalar is sufficient to stabilize the RG evolution of the scalar quartic couplings is explicitly shown in Eq.~(\ref{delta_lambda}). Through the scalar mixing, larger values for the SM-like quartic coupling $\lambda_1$ are needed to recover the same physical mass. Conversely, the heavy Higgs boson $h_2$ contributes additively to the EW boundary condition of $\lambda_1$, proportionally to the scalar mixing angle. This means that to obtain the same ``uplifting'' of the coupling to the stability region, larger $h_2$ masses are required when considering smaller mixing angles $\alpha$ (per fixed vev $x$). Second, for a fix value of $x$, $m_{h_2}$ cannot grow indefinitely to compensate for smaller and smaller values of $\alpha$, being limited by the occurrence of a Landau pole for $\lambda_2$.
Third, positive and negative values for $\alpha$ (and similarly for $\lambda_3$ in our conventions) are not symmetric, as clear from Eq.~(\ref{RGE_lambda3}). In fact, the running of $\lambda_3$ via said equation depends on the {\it relative} sign between ($\lambda_1 + \lambda_2$) and $\lambda_3$. Given that $\lambda_1,\, \lambda_2 > 0$, there will be constructive$\backslash$destructive interference when $\alpha$ assumes positive$\backslash$negative values. This means that the evolution of $\lambda_3(Q)$ is more stable for negative angles, that hence appear to be favored with respect to positive angles. Interestingly, negative values are physically preferred in some cases if one tries to explain the enhancement in the diphoton signal of the SM-like Higgs boson~\cite{Basso:2012nh}. Finally, heavy neutrinos can play a role in destabilizing the evolution of $\lambda_2$, as the top-quark for the SM quartic coupling, if heavy enough (see Eq.~(\ref{RGE_lambda2})). The request of stability of the model up to a certain scale then imposes an upper bound on the RH neutrino Yukawa couplings and therefore on the heavy neutrino masses (per fixed vev). Unless otherwise specified, we will consider $M_{\nu_h} = 200$ GeV throughout this Letter. 

\begin{figure}[!ht]
\centering
\includegraphics*[width=0.95\linewidth]{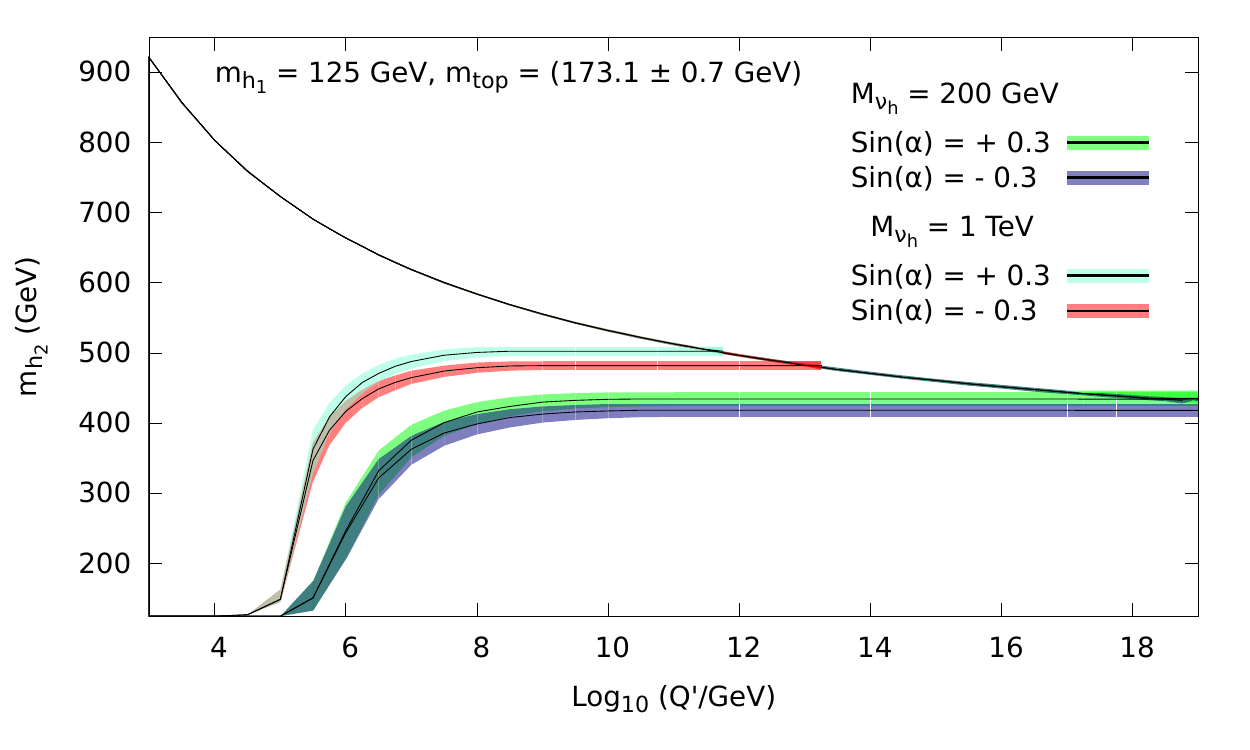} 
\caption{Allowed EW-scale heavy Higgs ($h_2$) masses as a function of the validity scale $Q'$ for $\sin{\alpha}=\pm 0.3$. Line thickness reflects a $1\sigma$ variation of the top-quark mass. Here, $x=7.5$ TeV. (Allowed masses are in the area inside the black lines).}
\label{fig:mh2_Qp_a03}
\end{figure}

\begin{figure}[!ht]
\centering
\includegraphics[width=0.95\linewidth]{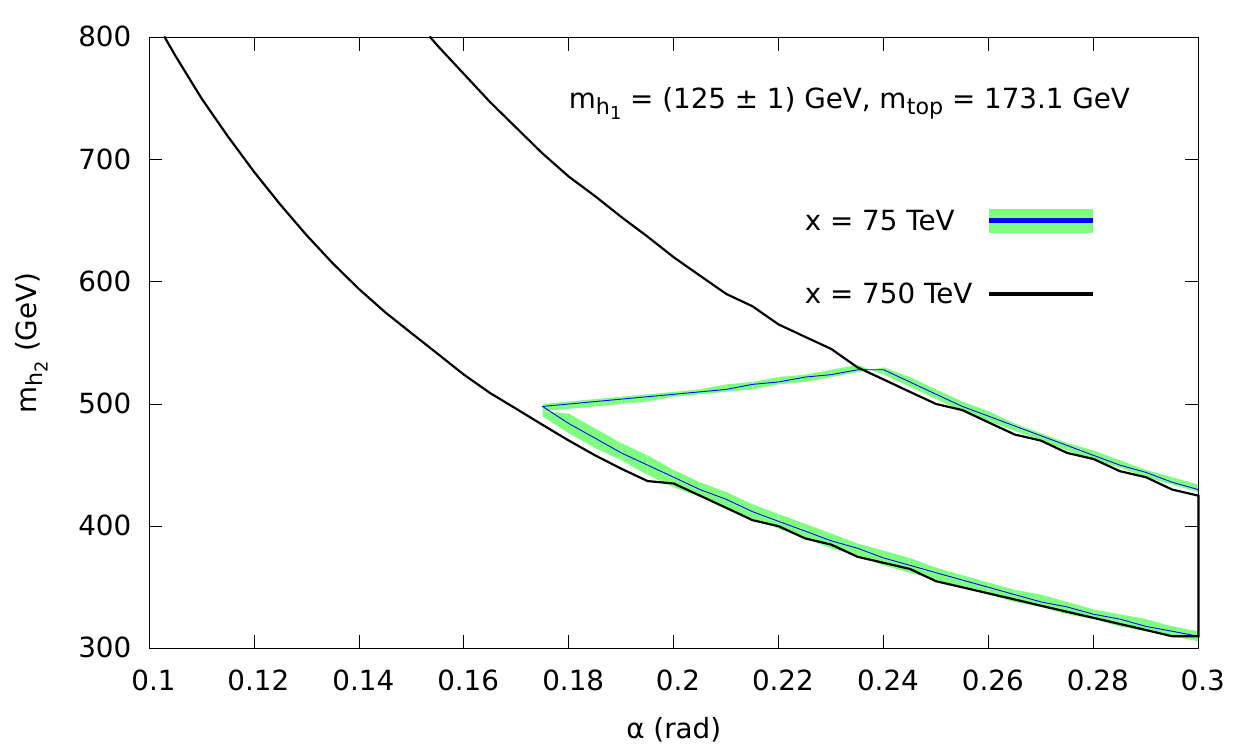} 
\caption{Allowed heavy Higgs mass as a function of the scalar mixing angle for {\it positive} values only. Thickness of the lines represent a $1\sigma$ variation on the SM-like Higgs mass. Here, $Q'=10^{19}$ GeV. (Allowed masses are in the area inside the lines).}
\label{fig:mh2_a_Qp19}
\end{figure}

\begin{figure}[!ht]
\centering
\includegraphics[width=0.95\linewidth]{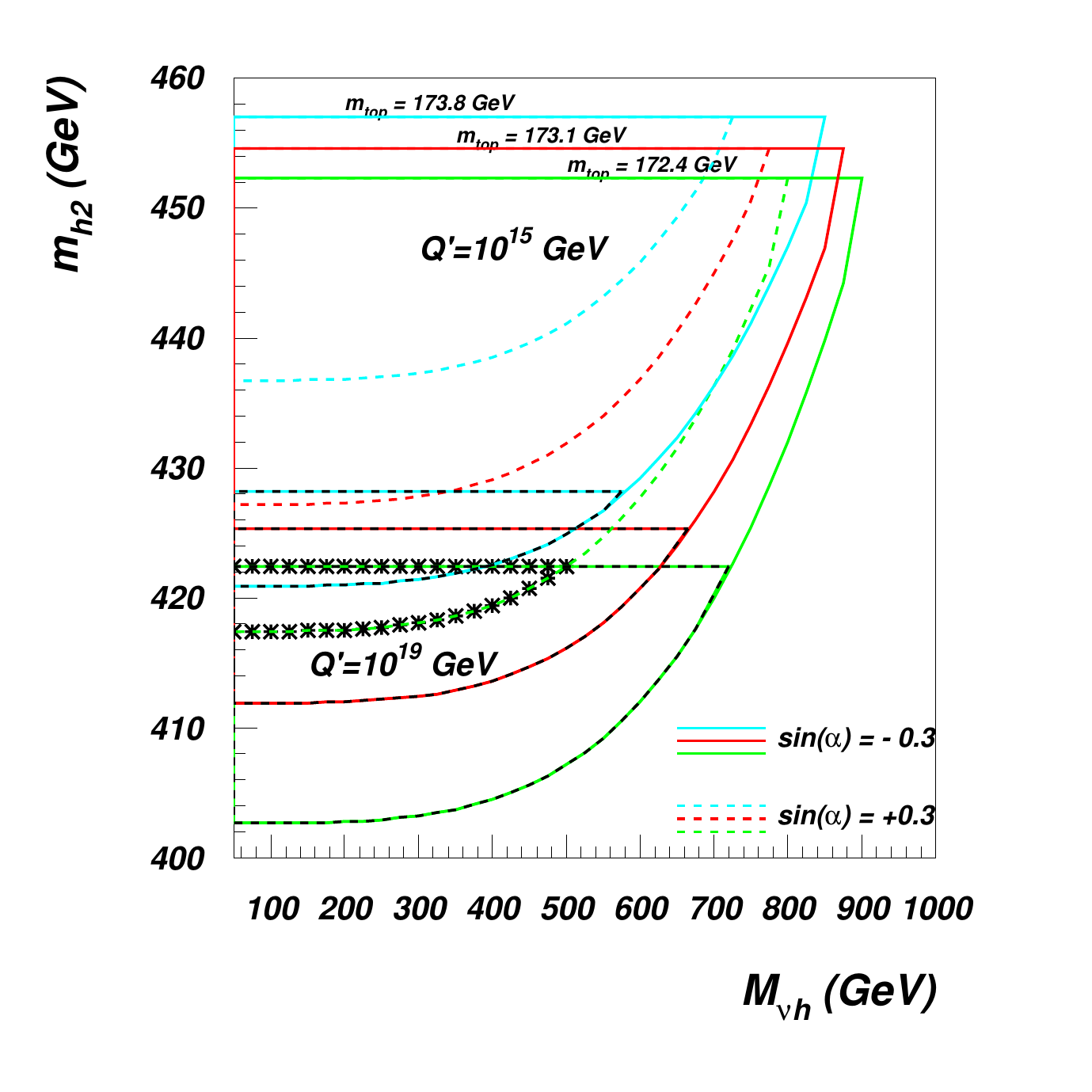} 
\caption{Allowed heavy Higgs masses as a function of the heavy neutrino masses. Color lines for $Q'=10^{15}$ GeV, dashed-colored-on-black lines for $Q'=10^{19}$ GeV; solid and dashed colored-black lines are for $\sin{\alpha}= - 0.3$, dashed-colored and dashed-colored-on-thick-black lines are for $\sin{\alpha}= 0.3$. Here, $m_{h_1} = 125$ GeV, $x=7.5$ TeV, and $\sin{\alpha}=0.3$. (Allowed masses are in the area inside the lines).}
\label{fig:mh2_vs_mhn_03}
\end{figure}

Fig.~\ref{fig:mh2_Qp_a03} shows most of these features at once, for $x=7.5$ TeV. There exist values for $m_{h_2}$ that are stable up to the Planck scale for negative scalar mixing angles for all the considered top-quark masses, but the similar case for positive values exist only for $m_{top} < 172.8$ GeV (i.e., for the central value of $m_{top}$, the model is not stable up to Planck scale for positive scalar mixing angle when the vev $x$ is small, see Fig.~\ref{fig:mh2_a_Qp19}). Further, it shows that if the RH-neutrino Yukawa couplings are large, the model can be stable only up to scales smaller than $M_{Pl}$, again with negative scalar mixing angles being stable for larger scales than the $\alpha >0$ case. We get that for $\sin{\alpha} = -0.3$, $400 < m_{h_2}/\mbox{GeV} < 430$ (see Fig.~\ref{fig:mh2_vs_mhn_03}), with Eq.~(\ref{delta_lambda}) telling that the allowed heavy Higgs mass scales with $\sin{\alpha}$ per fixed vev (i.e., for $\sin{\alpha} = 0.1$, $m_{h_2} \simeq 1.2$ TeV is stable up to $M_{Pl}$).

\begin{figure}[!ht]
\centering
\includegraphics[width=0.95\linewidth]{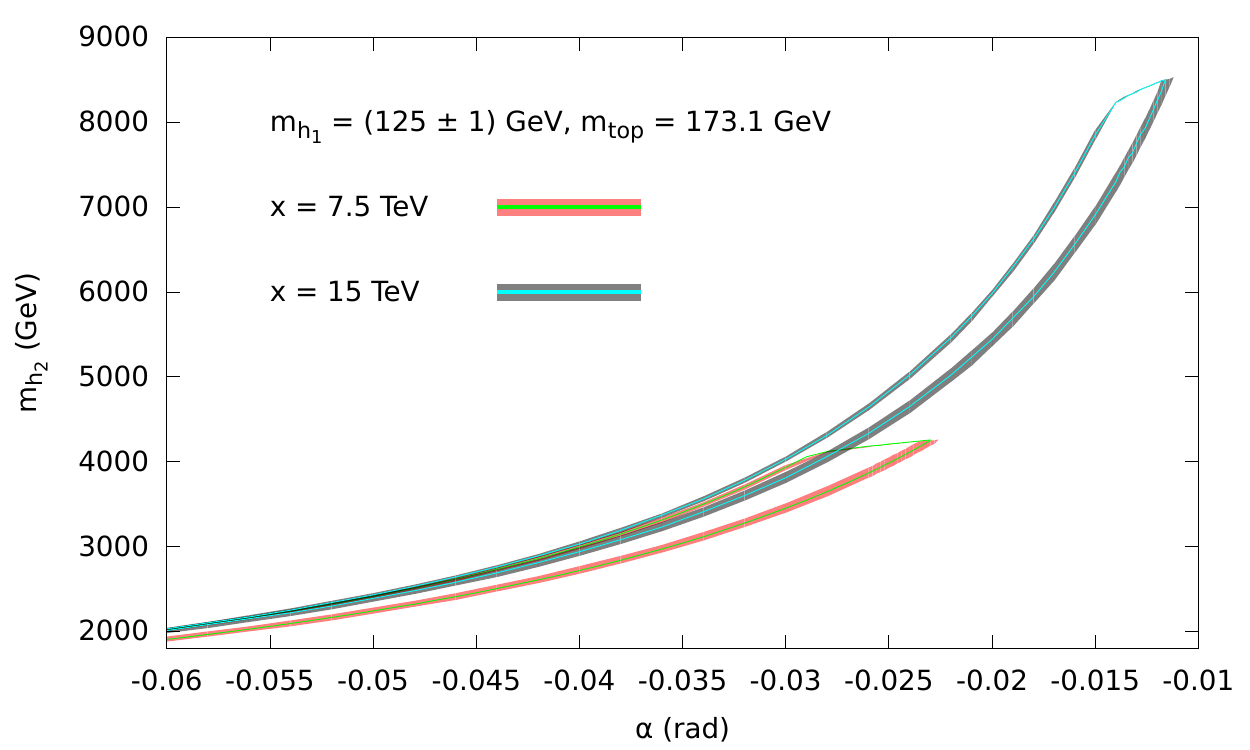} 
\caption{Allowed heavy Higgs mass as a function of the scalar mixing angle for {\it negative} values only. Thickness of the lines represent a $1\sigma$ variation on the SM-like Higgs mass. Here, $Q'=10^{19}$ GeV. (Allowed masses are in the area inside the lines).}
\label{fig:mh2m_am_Qp19}
\end{figure}

The scaling of $m_{h_2}$ with $\sin{\alpha}$ is displayed in Figs.~\ref{fig:mh2m_am_Qp19}--\ref{fig:mh2_a_Qp19}, for negative and positive angles, respectively. It is clear that for small $x$ values, negative angles are favored over positive values. In fact, one needs very large vev values $\mathcal{O}(100)$ TeV for $m_{h_2}$--$\sin{\alpha}$ combinations stable up to $M_{Pl}$ to exist, for positive angles smaller than $0.1$ rads. For negative angles, there exist heavy Higgs masses stable up to the Planck scale for $\sin{\alpha} \lesssim -0.025(-0.01)$ for $x=7.5(15)$ TeV. The substantial difference comes from the evolution of $\lambda_3$, that runs less when negative, such that the violation of vacuum stability is pushed to higher scales. For positive angles, it occurs instead much earlier, and larger vevs (per fixed $m_{h_2}$) are required to keep Eq.~(\ref{cond_2}) satisfied.
Depending on the ($B-L$)-breaking vev $x$, Figs.~\ref{fig:mh2m_am_Qp19}--\ref{fig:mh2_a_Qp19} show that a minimum scalar mixing angle exist (in absolute value), roughly equal to $-0.023(-0.011)$ rads for $x=7.5(15)$ TeV. In the positive domain, for $x=75$ TeV one needs  $\sin{\alpha} \geq 0.17$. For positive angles, a $\mathcal{O}(10\%)$ precision in the measurement of the Higgs cross sections is hence sufficient to test this model for $x<\mathcal{O}(100)$ TeV. The LHC is expected to reach this precision, and therefore can easily confirm/exclude this region of the parameter space. Larger vevs and/or negative $\alpha$ values (that allow for much smaller minimum angles) require higher precision and therefore can be tested only at future linear colliders.

Regarding the allowed heavy Higgs masses, they do not depend substantially on the vev $x$, because the condition that is restricting the parameter space is given by the triviality bound on $\lambda_1$, on which $m_{h_2}$ impinges via Eq.~(\ref{delta_lambda}), independent from $x$. However, the existence of a minimum scalar angle and its value depend on $x$. As remarked, one cannot grow $m_{h_2}$ indefinitely to compensate for smaller $\sin{\alpha}$ due to the appearance of a Landau pole for $\lambda_2(Q_{EW}) \sim \frac{m^2_{h_2}}{2 x^2}$. Interestingly, when considering negative angles, violations of vacuum stability appear earlier than the Landau pole for $\lambda_2$ when increasing the ($B-L$)-breaking vev $x$ to very large values (bigger than $\mathcal{O}(50)$ TeV), opposite to the previous observation. This is because $\lambda_3(Q_{EW}) \sim \sin{2\alpha}\frac{m^2_{h_2}}{x}$ (see Eq.~(\ref{inversion})), i.e., the boundary condition on $\lambda_2$ gets smaller than the one on $\lambda_3$ for increasing values of $x$, causing violation of the vacuum stability earlier in the running.

Fig.~\ref{fig:mh2_vs_mhn_03} shows that an upper limit on the heavy neutrino mass also exists for fixed $x$ and $\sin{\alpha}$ values, depending on the scale of validity of the model. This was clear already from Fig.~\ref{fig:mh2_Qp_a03} and from the behavior of the top-quark in the SM. In the case of the type-I seesaw mechanism under discussion, heavy neutrinos cannot be heavier than about $700(500)$ GeV when imposing stability up to the Planck scale, for $\sin{\alpha}=-0.3(+0.3)$. When $\sin{\alpha}=-0.1(-0.05)$, $M_{\nu_h} \lesssim 1(2)$ TeV. These values are for $x=7.5$ and scale naively with it. The plot shows also a strong sensitivity from the actual top-quark mass, and it reinforces the observation that positive scalar mixing angles are stable up to the Planck scale for small $x$ values only for top-quark masses smaller than its central value.

We conclude by noticing that, depending on the $Z'$ mass and couplings, heavy neutrinos can be discovered at the LHC (and possibly at future colliders) up to masses of $\mathcal{O}(1)$ TeV (see for example Ref.~\cite{pheno}), hence covering the available parameter space for $Q'=10^{19}$ GeV and $\sin{\alpha}=0.3$. In this case the heavy Higgs mass must lay around $420$ GeV, certainly within the range of the LHC at $\sqrt{s}=14$ TeV for the latter value of $\alpha$. Obviously, a heavier $h_2$ is required as the scalar mixing angle gets smaller, therefore decoupling the former. In this case, also the heavy neutrino can get heavier and hence more difficult to discover.

\vspace{5cm}

\section{Conclusions}
\label{sect:conclusions}
It is well established that the presence of an extra scalar allows for the so-enlarged SM to be valid up to the Planck scale. In this Letter we showed the case for the minimal $Z'$ models, taken as representative for this type of minimal SM extensions. It was shown in Ref.~\cite{Basso:2012nh} that negative scalar mixing angles can be employed to explain the SM-like Higgs-to-diphoton signal. The primary result of this work is that they are also favored for small values of the ($B-L$)-breaking vev with respect to positive values when stability up to the Planck scale is requested. It was also shown that a minimum value exist for the (absolute value of the) scalar mixing angle as well as a maximum value for the heavy neutrino mass, per fixed ($B-L$)-breaking vev values, and how they depend on the latter. Both these findings are important for the searches at the LHC (and at future colliders). On the one side, one does not need to reach infinite precision in the knowledge of the SM-like Higgs cross sections to rule out (or find hints for) SM extensions. On the other side, light TeV-scale particles are predicted. Despite the lack of extensive studies of their LHC discovery potential in this model, partial studies exist that confirm the possibility to detect them if not too heavy.

\section*{Acknowledgements}
I would like to thank G.~M.~Pruna for valuable discussions that triggered this work, and J.~J.~van~der~Bij for reading the manuscript. I also thank S.~Iso and F.~Staub for pointing out the error in Eq.~(\ref{RGE_lambda2}).
This work is supported by the 
Deutsche Forschungsgemeinschaft through the Research Training Group grant
GRK\,1102 \textit{Physics at Hadron Accelerators} and by the
Bundesministerium f\"ur Bildung und Forschung within the F\"orderschwerpunkt
\textit{Elementary Particle Physics}.

\bibliographystyle{elsarticle-harv}

\end{document}